\documentclass[runningheads]{llncs}
\usepackage{graphicx}
\usepackage{times}
\usepackage{xcolor}
\usepackage{soul}
\usepackage[utf8]{inputenc}
\usepackage[small]{caption}

\usepackage{algorithm}
\usepackage{algpseudocode}
\usepackage{subfig}
\usepackage[euler]{textgreek}
\usepackage{amssymb}
\usepackage{graphicx}
\usepackage{csquotes}
\usepackage{enumerate}
\usepackage{latexsym}
\usepackage{url}
\usepackage{subfig}
\usepackage{array}
\usepackage{spverbatim}
\usepackage{comment}
\usepackage{eurosym}
\usepackage{booktabs}
\usepackage{adjustbox}
\usepackage{subfig}
\usepackage{svg}
\usepackage{multirow}
\usepackage{adjustbox}
\usepackage{xcolor}
\begin{document}
\title{A Human-Machine Collaboration Framework for the Development of Schemas}

\author{Nicos Isaak\inst{1}\orcidID{0000-0003-2353-2192}}
\authorrunning{N. Isaak}

\institute{Computational Cognition Lab, Cyprus\\
	\email{nicosi@acm.org} }

\maketitle

\begin{abstract}
The Winograd Schema Challenge (WSC), a seemingly well-thought-out test for machine intelligence, has been proposed to shed light on developing systems that exhibit human behavior. Since its introduction, it aimed to pivot the focus of the AI community from the technology to the science of AI. While common and trivial for humans, studies show that it is still challenging for machines, especially when they have to deal with novel schemas, that is, well-designed sentences that require the resolving of definite pronouns. As researchers have become increasingly interested in the challenge itself, this presumably necessitates the availability of an extensive collection of Winograd schemas, which goes beyond what human experts can reasonably develop themselves, especially after proposed ways of utilizing them as novel forms of CAPTCHAs.

To address this necessity, we propose a novel framework that explicitly focuses on how humans and machines can collaborate as teammates to design novel schemas from scratch. This is being accomplished by combining two recent studies from the literature: i) Winventor, a machine-driven approach for the development of large amounts of Winograd schemas, albeit not of high quality, and ii) WinoFlexi, an online crowdsourcing system that allows crowd workers to develop a limited number of schemas often of similar quality to that of experts. Our proposal crafts a new road map toward developing a novel collaborative platform that amplifies human and machine intelligence by combining their complementary strengths. 

\end{abstract}
\section{Introduction}
The Winograd schema challenge (WSC) \cite{levesque2012winograd}, the task of resolving definite pronouns in carefully constructed sentences, has been proposed to help researchers build systems that understand human behavior \cite{levesque2014our}. Broadly speaking, this challenge is about resolving sentence ambiguities because the information needed to resolve definite pronouns is not grammatically present. It is believed that the challenge will play a significant role in the development of a wide swath of AI applications as a step towards the development of machines that will automate or enhance basic human abilities, a traditional AI goal laid back in the late 50s \cite{kn:michael2013machines}.

Scholars seem to agree that, while strictly following the challenge rules, the WSC is relatively trivial for humans and tricky for machines \cite{kn:bender2015establishing,morgenstern2016planning}, especially when knowing that opaque-statistical techniques can teach us nothing about human behavior itself \cite{marcus2019rebooting}. 

However, along a completely different tangent, when approaching the challenge as another pronoun resolution problem, there are subsets of schemas that can be solved with high accuracy by utilizing clever tricks that, sometimes, even humans can hardly determine \cite{KOCIJAN2023103971}. In this sense, when completely removing the science of AI factor, which propels us to figure out how humans analyze Winograd schemas to come to conclusions about events and participants, and by focusing on the technology of AI, we can build narrow savants that show but not exhibit real commonsense and reasoning abilities when tackling Winograd schemas \cite{icaart22}.

Especially with the advent of large language models, many researchers have pursued a course of achieving high accuracy results in specialized subsets of schemas, often comparable to those of humans, but with ways that have nothing to tell us about human behavior itself \cite{icaart22}. It is well-known that these systems, although they can achieve high accuracy results in specified subsets of schemas by analyzing the pattern of words, have nothing to do with human results, as they often fail to answer with scores even below chance \cite{KOCIJAN2023103971}. 

To put it succinctly, strictly following the purpose of the challenge, that is, focusing on the science and not the technology of AI, could lead us towards developing machines with commonsense and reasoning abilities similar to those found in humans. This is well-stated and analyzed in another paper of ours, where we shed light on how to build commonsense and reasoning systems that analyze and tackle schemas the way humans do \cite{icaart22}.

At the same time, this evidence points to the need for more research in identifying ways to develop intelligent machines. In this line of research, in order to bring more AI research into the field, recent work has demonstrated the possibility of using the WSC as a novel form of CAPTCHA \cite{GCAI-2018:Using_Winograd_Schema_Challenge}. Although this approach is interesting, it suffers from the lack of available Winograd schemas due to schema development difficulties \cite{morgenstern2016planning}. In either case, for future developed systems, it seems that the availability of continuously replenished Winograd schemas is necessary for further work and progress.

Following this line of research, two recent works have demonstrated the possibility of developing Winograd schemas from scratch \cite{10.1007/978-3-030-35288-2_24,icaart20}. On the one hand, Winventor \cite{icaart20}, a machine-driven approach, can automate the schema development process and considerably help experts in the development task. On the other hand, WinoFlexi \cite{10.1007/978-3-030-35288-2_24,icaart20}, a flexible crowdsourcing approach, allows members of crowdsourcing platforms to collaborate exclusively for the development of Winograd schemas. While Winventor can produce large amounts of schemas, their quality is limited because of various peculiarities in sentences collected from the WWW. On the contrary, WinoFlexi can supply us with a limited number of schemas of high quality, similar to those developed by experts.

Aiming to develop a new system that can amplify human and machine intelligence via the development of vast amounts of high-quality schemas, this work proposes a new Human-Machine Collaboration framework-- for simplicity, we call it WinoFusion. WinoFusion is a novel hybrid approach that combines the strengths of Winventor \cite{icaart20} and WinoFlexi \cite{10.1007/978-3-030-35288-2_24} into one powerful collaborative platform aiming to provide the crowd the necessary tools to design vast amounts of schemas, in the shortest time possible. Via WinoFusion, we can get the best of both worlds based on which the crowd can enhance their creativity and motivation to develop schemas.

We start by offering some guidelines on the challenge and continue with our related work and motivation section. Then, we present our Human-Machine Collaboration framework and conclude with our conclusion and future work section.

\section{Challenge Basics}
The Winograd Schema Challenge (WSC) aims to build machines with basic human abilities \cite{levesque2012winograd,levesque2014our}. The challenge consists of pairs of halves, and the objective is to resolve a definite pronoun in each half. Each half comprises a sentence, a question, and two possible pronoun targets or answers. Moreover, in each half, there is a special word that, when replaced, the answer also changes. It is believed that systems able to tackle the challenge would presumably support a wide range of commonsense and reasoning abilities that would help in the development of several other AI applications. The following WSC schema\footnote{Created by Workers on WinoFlexi \cite{10.1007/978-3-030-35288-2_24}} illustrates an example: 
\subsubitem First-half: \emph{Sentence}: The [martial artist]\textsuperscript{1st pronoun-target} defended himself from the [drag dealer]\textsuperscript{2nd pronoun-target} because [he]\textsuperscript{define-pronoun} was [violent]\textsuperscript{special-word}. \emph{Question}: Who was violent? \emph{Answers}: The drug dealer, the martial artist. \emph{Correct-Answer}: The drug dealer. 
\subsubitem Second-half: \emph{Sentence}: The [martial artist]\textsuperscript{1st pronoun-target} defended himself from the [drag dealer]\textsuperscript{2nd pronoun-target} because [he]\textsuperscript{define-pronoun} was [under-attack]\textsuperscript{special-word}. \emph{Question}: Who was under-attack? \emph{Answers}: The drug dealer, the martial artist. \emph{Correct-Answer}: The martial artist.

For the past eight years, there has been a rapid rise in the designing of narrow opaque systems to tackle the challenge, albeit without exactly reaching human adult results. On the one hand, English-speaking adults have no difficulty with the challenge as they achieve accuracy results up to 100\% almost in every type of schema, while, on the other hand, currently developed systems only achieve high accuracy results up to 90\% on narrowly specified subsets of schemas. In this sense, for the foreseeable future, we cannot claim to reach human-level abilities until AI systems achieve accuracy results up to 100\% on every type of schema. 

In short, we can approach the challenge in two ways, either as the means towards achieving human-level intelligence in machines or as another challenge we must tackle to improve our benchmark leaderboard position. In case we choose the latter, there is not much we can do, considering state-of-the-art approaches that achieve high accuracy results, all pinpointing to its defeat, at least for a specialized subset of schemas \cite{KOCIJAN2023103971}. However, in the case of the former point of view, there are still many stepping stones we need to cross towards developing machines with basic human-level abilities.

\section{Related Work - Motivation}
According to the literature, the development of new schemas by experts is a laborious task requiring creativity, inspiration, and motivation \cite{morgenstern2016planning}, and perhaps, not unsurprisingly, there seem to exist a limited number of WSC datasets that have been widely used \cite{KOCIJAN2023103971,levesque2012winograd,kn:Rahman:2012:RCC:2390948.2391032}. 

It seems that the development of new schemas from scratch deserves more research attention, given that the availability of Winograd schemas seems disproportional to their demand and potential impact. Perhaps not unrelated to the limited availability of schemas, the first and only WSC took place in 2016 as a side event of IJCAI, where contestants did not manage to earn the price of \$25000 \cite{morgenstern2016planning}. According to the challenge organizers, the challenge itself was found to be too troublesome and difficult to handle at regular intervals. In this line of work, the WSC is now part of the GLUE-benchmark \footnote{\url{https://gluebenchmark.com}} where new Winograd schemas could potentially enhance the language understanding evaluation task. Of course, in retrospect, this led to the development of various approaches that, although they tackled subsets of schemas with high accuracy results, were often done without considering the challenge purpose as the means to understanding the innards of basic human behavior while disambiguating pronouns. In hindsight, when the challenge was introduced, it was not easy to consider the possibilities of several obvious lines of attack by narrow statistical and opaque approaches such as neural network ones. On the other hand, it is not feasible even for humans to determine how machines could find word patterns that could lead them to the successful resolution of pronouns in schemas.

In a relevant study \cite{GCAI-2018:Using_Winograd_Schema_Challenge}, we showed that the challenge can form a novel form of CAPTCHAs (Completely Automated Public Turing test to tell Computers and Humans Apart) in an effort of bringing security researchers to the field. In this regard, a new schema replenishment mechanism was necessary to display schemas to identify humans from bots. Reportedly, since then, few studies have investigated how we can develop Winograd schemas from scratch, like the Winventor and WinoFlexi systems \cite{10.1007/978-3-030-35288-2_24,icaart20}.

Winventor \cite{icaart20} is a system that automatically develops schemas from scratch and considerably helps humans in the development task. In its simplest form, it uses several components that enhance the schema development process applied to sentences found on the English Wikipedia. If it cannot develop a schema, it only develops a half ---the first instance of a schema. Broadly speaking, it parses each examined sentence to select sentences that include pronouns and more than two nouns/proper nouns. Next, it parses each examined sentence via a spelling-correction component to fix possible abbreviations, spelling errors, or misspellings of words. With the help of the spaCy\footnote{\url{https://spacy.io}} dependency parser, it builds semantic relations, which show how the sentence's nouns or proper nouns relate to the sentence's pronouns. Through another component, it selects the best possible pronoun-target pair, either two nouns or proper nouns with the same number and gender. The next step is developing the first half's question through a specified question generator \cite{heilman2009question}. The last step includes finding and substituting the first half's special word to build the second half.

WinoFlexi \cite{10.1007/978-3-030-35288-2_24} is an online collaboration platform that enhances the schema development process, where crowdsourced workers collaborate to design schemas in real-time. Specifically, WinoFlexi brings together workers from various geographic locations \cite{christoforaki2014step} to work on the development of Winograd schemas. To that end, it incorporates several mechanisms that guide humans with no prior knowledge in designing new schemas of similar quality to most typical existing collections. Among others, WinoFlexi's mechanisms include a training session that familiarizes workers with the schema development process by asking them to correctly resolve randomly selected schemas designed by experts. To do that, it categorizes workers into two groups: Contributors and Evaluators. Contributors are workers who develop schemas, that is, pair of halves consisting of sentences, questions, and answers. On the other hand, the Evaluators are qualified Contributors who validate schemas. At the same time, the engine is cheat-proof, and it uses quality controls that benefit non-dubious workers and ban dubious ones. To illustrate, it uses quality controls, namely, test questions, the ban score, and the schema hardness score. The test-question approach is based on the adaptive interjection of questions during the schema development process and evaluation, where workers who correctly resolve them get a positive score. The ban score, a specified way to deal with cheating, automatically bans workers with a sufficiently predefined low score.

The schema hardness mechanism is WinoFlexi's way of giving Contributors feedback regarding their developed schemas' structure. This is a metric that labels each schema with a hardness score that indirectly shows if a schema is considered hard or easy in relation to their already developed schemas --- e.g., if the majority of contributor schemas are easy to solve, WinoFlexi prompts them to develop more complex schemas. Experiments showed that WinoFlexi experiments ran for one week and yielded more than 165 schemas from 50 workers aged 18 to 65. However, it seems the engine could not motivate them to develop large amounts of schemas. Specifically, during the experiments, it was found that only 50 workers decided to participate, where the highest score was 250 points and the lowest -70. Additionally, the mean response time across all workers was 1.48 minutes, and the average time for the best worker was 1.66 minutes. Also, it was found that 60\% of bonuses were offered to the top five workers.

It seems that a novel Human-Machine Collaboration approach that blends the advantages of the studies mentioned above would enhance the schema development process. In this regard, inspired by having humans and machines acting as associates rather than supervisors and tools, this work introduces WinoFusion, which brings the schema development process into a new era, where both humans and machines are evolving from supervisors and assistants to associates and collaborators \cite{de2018automation}. According to the literature, Human-Machine Cooperation (HCM) \cite{doi:10.1080/001401300409044} is necessary to introduce new stakes, which were introduced a long time before with Human-Computer Interaction (HCI). On the one hand, machines could provide autonomous agents for different tasks, and on the other hand, humans could control the whole process with the optimum goal of producing efficacious results. \cite{10.1145/3301275.3302324,doi:10.1080/001401300409044}. 

To put it succinctly, by amplifying human and machine intelligence, we can address some of their weaknesses \cite{vandenhof2019hybrid}. This is why Human-Computer Interaction has moved to the collaboration era between humans and machines \cite{kennedy2018imperative}. In essence, our proposed framework utilizes the Winventor replenishment mechanism to inspire and motivate crowdworkers to be creative based on a novel interaction that amplifies human and machine intelligence by combining their complementary strengths. The design of appropriate crowdsourcing mechanisms for our particular task is the focus of the rest of this paper.

\section{WinoFlexi: A Human-Machine Collaboration Framework}
Aiming to develop systems that amplify human and machine intelligence by developing vast amounts of high-quality schemas, this work presents WinoFusion. This new hybrid framework is bringing together the strengths of Winventor \cite{icaart20} and WinoFlexi \cite{10.1007/978-3-030-35288-2_24} into one powerful, flexible platform able to provide the crowd with the necessary tools to design vast amounts of schemas, in the shortest time possible. Through WinoFusion, the crowd can enhance their creativity and motivation to develop quality schemas that were initially designed by Winventor. In the following lines, we will underline the major components of our framework by explaining how it works and handles the schema development process. Consequently, we will show how it blends WinoFlexi and Winventor mechanisms into a single novel collaboration engine.

\subsection{Qualificators \& Supervisors}
WinoFusion groups workers into two predefined categories, namely Qualificators and Supervisors. Qualificators are workers who modify schemas initially developed by Winventor. On the other hand, Supervisors oversee the schema design process by evaluating schemas modified by Qualificators. To illustrate, When a Qualificator accepts to modify a schema, WinoFusion checks if the schema parts are related. Like in WinoFusion \cite{10.1007/978-3-030-35288-2_24}, this is done with the relatedness heuristic approach that checks if the parts relate through common words. Additionally, given that Winventor cannot select the correct answer to each schema half, WinoFusion checks if the Qualificators have selected the correct answers. As with WinoFLexi's Contributors, for Qualificators to become Supervisors, they need to have both a high amount of correctly validated schemas and a score bigger than a specified threshold ---each valid schema increases a contributor's score, and each invalid schema decreases it. Moreover, in addition to WinoFlexi's similarity tools that mark schemas that follow a similar pattern, the engine notifies Supervisors of specialized factors like correctness and engagement through schema analytics based on word-level analysis. Correctness refers to stats relating to grammar mistakes per schema so far and engagement to a simplified measure regarding each Qualificator's participation in the schema design process.

\subsection{Advanced Collaboration Techniques}
Through better evaluation controls, WinoFusion adds to the WinoFlexi mechanisms to enhance quality and improve worker's experience. Via advanced collaboration techniques, it takes advantage of the variability that generally stems from Winventor's schemas and the ``wisdom of the crowd" \cite{Simoiu2019ALS}, which is arguably better than any individual worker's opinion. 

\subsubsection{Continuously-Replenished Mechanism:}
The main feature of our proposed framework is a continuously replenished mechanism that provides the crowd with schemas. This mechanism is a two-way procedure that considerably helps Qualificators develop more schemas by enhancing their motivation and inspiration abilities. At the same time, it works as a Winventor feedback mechanism that leverages the schema development process as accurately as possible. Compared to WinoFlexi, where the crowd had to develop schemas from scratch without help, WinoFusion displays a list of available schemas that can be easily evaluated. The list, which displays only a subset of the schemas, allows each worker to select a schema at a time, requiring them to provide immediate feedback before selecting another one. With the schema selection process, all the Qualificator form fields are automatically filled out, meaning that, for each half, there is a sentence, a question, and two possible pronoun targets. Moreover, within several options, each Qualificator needs to provide immediate feedback regarding each schema's quality by selecting the correct answer to a predefined list of yes/no questions:
\subsubitem{1.) This is a valid schema, and I do not need to make any changes: yes/no}
\subsubitem{2.) This was not a valid schema, but I could modify it to a valid one: yes/no}

\noindent
However, in case of choosing to completely discard a schema, the engine automatically resets and proposes a list of two other yes/no questions:
\subsubitem{1.) This is not a valid schema, and I cannot modify it to a valid one: yes/no}
\subsubitem{2.) This is not a valid schema, and although I can modify it, I do not like its subject: yes/no}

\subsubsection{Schema Qualification Mechanism:}
Within WinoFusion mechanisms, we want to make sure that both quality schemas are correctly qualified and that a plethora of schemas are created from each Winventor example. At the same time, given that every schema can be examined by multiple Qualificators, we do not want to overburden our workers with the development of numerous schemas stemming from the same schema template. To avoid these kinds of problems, WinoFusion imposes specified restrictions on the number of times a specified example is used. Compared to WinoFlexi, within WinoFusion, every examined schema is evaluated by at least three Qualificators through an automatic procedure that takes place on specific days of the Weekend, set by the super administrator ---more on the administrator's role can be found in the original WinoFlexi paper. Nevertheless, this evaluation mechanism adds to the quality of the developed schemas because of the "wisdom-of-the-crowd" factor \cite{Simoiu2019ALS}. After that, each schema is evaluated by our Supervisors and accordingly flagged as valid-finished or valid-pending. Valid-finished means that our schema template is removed from our Qualificator-proposed list while valid-pending that more schemas can be derived from it.

\subsubsection{Semi-Schema Templates:} 
The idea is to provide the crowd with sentences that meet certain criteria rather than solely schemas designed by Winventor. Winventor analysis indicates that in its first automated evaluation experiment, a significant fraction of the examined sentences were rejected because of pure parsing \cite{icaart20,isaak2021blending}. Since semantic relations are challenging to develop via pure NLP \cite{schubert2015semantic}, there is a big chance that many rejected sentences could be considered valid under the crowd evaluation process.

To that end, there is a big chance that the crowd could utilize these sentences to develop new schemas. To handle problems like this, WinoFusion allows Wikipedia sentences with at least two nouns of the same gender and number to be displayed for further analysis. Without eliminating their apparent shortcomings, under proper design modifications, these semi-schema templates might synergize the development of quality and diverse sets of schemas. However, to dodge the possibility of having our Qualificators working with hard cases of semi-schema templates, their displayed factor should be set as low as 10\% as in WinoFlexi's test-question mechanism \cite{10.1007/978-3-030-35288-2_24}.

\subsection{Enhanced Quality Controls}
Since dealing with cheating is a major challenge \cite{10.1145/3148148,10.1007/978-3-030-35288-2_24}, the following mechanisms add to the WinoFlexi's quality control responsible for the quality of the developed schemas. As within WinoFlexi, we argue that crowdsourcing platforms should utilize tests as a method of assessment to verify that given workers indeed hold particular skills. In this sense, certain utilities, like training and test questions, lead to the integration of interactive mechanisms that motivate participants' engagement in the schema development process in a more careful way \cite{christoforaki2014step,hirth2011anatomy,peer2015beyond}. 

\subsubsection{Advanced Training:}
As with WinoFlexi, workers who want to register must pass a training phase to familiarize them with the schema development process. Compared to WinoFlexi, in the training phase, WinoFusion requires workers to correctly resolve randomly selected schemas and validate others manually developed by us. These schemas contain grammar mistakes, misspellings of words, and some word ordering problems. Given that prior experiments showed that motivation and inspiration play an important role in the schema development process, we want to acquire workers with both characteristics in their assets, meaning that workers who fail to develop schemas cannot proceed further. Finally, similarly to WinoFlexi, the length of the training phase for every newly registered Qualificator is automatically determined by how much the number of invalid schemas produced so far exceeds the number of valid ones.

\subsubsection{Enhanced Test-Question Control:}
As this is a challenging task that requires inspiration and creativity, we want to keep indubious Qualificators and ban dubious ones. One simple mechanism to achieve this is to use test questions during the schema development process. Test Questions refer to questions displayed by WinoFusion to Qualificators and Supervisors \cite{10.1007/978-3-030-35288-2_24}. In its simplest form, this assessment method verifies that our workers remember what a schema is and what the WSC is all about. Like in WinoFlexi, i) our workers are rewarded with a positive score for successfully resolving a Winograd schema, and ii) a test question has a $10\%$ probability of being displayed after every login. In addition to the adaptive interjection of test questions, WinoFusion randomly displays schemas that need to be approved/disapproved by both Qualificators and Supervisors. WinoFusion acquires validated schemas from Rahman and Ng's dataset \cite{kn:Rahman:2012:RCC:2390948.2391032} and unvalidated schemas from Winventor's database \cite{icaart20}. All in all, these additional test mechanisms might divert any dubious workers away from utilizing WinoFusion's mechanisms.

\subsubsection{Advanced Hardness-Metric Tool:}
Compared to WinoFlexi, WinoFusion leverages an advanced hardness-metric tool \cite{GCAI-2020:Winoreg} as a feedback mechanism for our Qualificators and Supervisors. This metric tool is based on a machine learning approach that outputs the hardness of a schema 300 times faster than WinoFlexi's mechanism. Via this hardness-metric tool, WinoFusion can provide immediate feedback during the schema design process. Moreover, within this tool, which can estimate the hardness index of any developed schema, WinoFusion aims to balance the quality of our developed schemas. To that end, if the majority of a Qualificator's schemas are considered easy to solve, then WinoFusion prompts them to develop schemas that are harder to solve.

\subsection{Motivation \& Inspiration Factor}
Recognizing that the schema development process is tedious and troublesome, \textit{WinoFusion} acts as a compelling impetus that motivates workers. Not pursuing such a course may be as mindlessly walking to a completely different path. The whole idea behind WinoFusion is to provide the crowd with the necessary tools to develop quality and diverse schemas based on various subjects. As previously stated, Winventor \cite{icaart22,isaak2021blending} can motivate and inspire experts to develop rich and diverse schema halves. In this regard, previous experiments showed that human experts were able to develop schemas for a variety set of sentence types. On the other hand, it was also shown that selecting the best schemas was challenging and troublesome. To address some of Winventor's weaknesses and, at the same time, to enhance its strength, WinoFusion provides a small arsenal of incentive tools.

\subsubsection{Bonus Banner:}
Previous research showed that workers need bonuses for high-quality results \cite{hirth2011anatomy,10.1007/978-3-030-35288-2_24}. In this sense, bonuses and other financial incentives can motivate workers to stay in place. Here, WinoFusion is taking the bonus factor to another level by showing the Qualificators the current amounts given to the crowd through a bonus banner. As other research showed \cite{Simoiu2019ALS}, these incorporated gamification aspects can encourage and motivate workers to keep themselves busy.

\subsubsection{Comments Banner:}
The ultimate goal is for WinoFusion to act as a collaborative task wherein multiple workers work together. The futility of this idea is to use the crowd's wisdom factor as it performs better than any average individual \cite{Simoiu2019ALS}. Participants who receive information about other people's choices behave better, leading to better results. Moreover, this would lead to the counter-effect of turning personal views closer to unbiased aggregate views \cite{madirolas2015improving}. For assessing productivity, WinoFusion displays a comment banner where each Qualificator can view and add comments regarding the schema development process.

\subsubsection{WinoFusion Inner-Workings:}
Focusing on explainable AI (XAI) and building systems that humans understand is essential in our days \cite{10.1145/3290605.3300831,de2018automation}. Considering the above, the idea of shedding light on how Winventor, WinoFlexi, and WinoFusion work to handle semantics to build schemas might play an essential role in the quality of the developed schemas. Moreover, people might be more likely to participate in research if they know that there is a machine behind it instead of humans. To that end, WinoFusion provides the crowd with an interactive tutorial on collecting sentences to build schemas and the necessary mechanisms that contribute to their enhancement. In this sense, all the information about our framework's inner workings will create a more trusting relationship between humans and machines to improve human productivity.

\subsubsection{Schema-Analysis Mechanism:}
Motivating people in various ways with encouragement mechanisms is essential for delivering quality results \cite{10.1145/3290605.3300831}. In this sense, different tools of encouragement are necessary for creating enjoyable working or training conditions \cite{evans1997tools}. For instance, research has shown that people spend more money when dealing with anthropomorphic slot machines \cite{riva2015humanizing}. As this approach might add to the schema quality, WinoFusion, upon each schema building, uses various encouragement messages that indirectly behoove Qualificators to focus on developing quality schemas. In this regard, and for every developed schema, WinoFusion displays statistical results, such as the different words they use to modify the schema, which are related to the part of speech of each word they use. At each time, WinoFusion displays stats regarding the differences between their modified version and the original schema developed by Winventor. Moreover, if the crowd is not using rich lexical representations, the engine prompts them with messages that nudge them to improve their writing style.

\subsubsection{Schema-Adaptation Mechanism:}
Should Qualificators have the best schemas to qualify, that will save us money and time, and most importantly, it will enhance Qualificator's productivity \cite{10.1145/3290605.3300831}. Thus, WinoFusion selects and rearranges the best schemas to display based on several factors, such as each pronoun target appropriateness, initially selected by Winventor based on the following order:
\begin{itemize}
	\item Schemas with the same number, gender, and pronoun-gender agreement.
	\item Schemas where the pronoun targets participate in triple relations.
	\item Schemas whose pronoun targets have the best Mitkov score values.
\end{itemize} 

Knowing that previous research showed that each schema sentence length is already in full swing with each imposed difficulties in resolving their pronoun targets \cite{GCAI-2020:Winoreg}, and that when dealing with machines and humans, handling biases is very crucial for having quality results \cite{10.1145/3290605.3300831}, we would need to employ a novel unbiased generalization language \cite{mitchell1980need} that will help towards the employment of quality schemas. 

For example, we know that children's guesses regarding new words' meanings depend on the structure of the language they are learning \cite{gathercole1997word}. To deal with the above, WinoFusion gathers crowd collective knowledge to characterize each Winventor schema as biased/unbiased. Through an on-the-fly machine learning technique, it automatically rearranges the schema selection process by promoting unbiased schemas on top of a suggestion list. 

\subsection{Winventor Adaptivity}
Recall that the ultimate goal is to blend and propel WinoFusion and Winventor work under a novel collaboration framework that will benefit humans and machines in producing quality schemas in the shortest possible time. Hence, from this point forward, it is crucial to show how Winventor adapts and modifies its inner workings to maximize its results based on the WinoFusion framework. In any different case, there would be a relative void in this emerging area on how Winventor benefits and propels the schema development process under the WinoFusion framework.

As other works have stated, a fundamentally different approach to enhance human and machine collaboration is crucial \cite{de2018automation}. In this sense, both humans and machines should elaborate to keep getting calibrated in the new ear where they should act as teammates. Recall that WinoFusion automatically analyzes the crowd schema selection process regarding the above factors and presents them to the crowd. At the same time, WinoFusion sends this to Winventor, which, on his side, enhances its schema development mechanism. According to the human-collected knowledge from WinoFusion, Winventor adapts and changes the way it works. More concretely, Winventor re-casts the whole process from just an ``interaction" to a ``collaborative relationship" \cite{de2018automation}. To that end, and through human guidance, Winventor is tasked to offer better solutions, leading to the development of quality schemas in the shortest time possible \cite{10.1145/3301275.3302324}.

WinoFusion analyzes its results based on how many schemas were developed from Winventor schema templates and how many, developed initially from Wikipedia sentences, were disqualified. Specifically, it analyzes how many schemas were developed regarding the pronoun target factors, namely the Markov score, gender, pronoun-gender, and number agreement. Also, it saves stats related to the subjects of the utilized schema templates. All these stats, which are crucial for enhancing the automation process, are sent to Winventor to change its inner workings. Given that originally, Winventor had limited capabilities or concern for direct human communication, we consider this feedback mechanism advantageous \cite{icaart20,de2018automation}.

In general, Winventor should analyze and compare the structure of valid schemas to invalid ones to determine their structure, such as each sentence's number of nouns and pronouns. The goal is to pick the most promising ones, leading to the development of quality schemas when subsequently searching and selecting sentences from the English Wikipedia. Within this regard, WinoFusion feedback could automatically update Winventor's internal working parameters, including but not limited to the acquired sentence-length and factor-priority list, which relate to how schemas are returned to Qualificators. For instance, the \textit{sentence-length} parameter, which limits each Wikipedia-acquired sentence length in words to a preferable size, indirectly thwarts long and complex sentences requiring high processing power to be utilized in the development of schemas. Moreover, the \textit{factor-priority} list parameter relates to how the schemas are returned to Qualificators according to mitcov, gender, number score, and each sentence's nominal subjects. Recall that WinoFusion stores, in real-time, the number of schemas that are correctly validated as such by our Supervisors. According to these results, the factor weight list is changing, and once a change is made, the priority list factor changes the way Winventor delivers its schemas.

\section{Conclusion and Future Work}
We have shown a novel human-computer interaction approach, where humans and machines evolve from tools and supervisors to teammates. Knowing that both humans and machines come with their own pros and cons, we have set the path towards building a new kind of collaboration platform that helps workers and machines design quality Winograd schemas in the shortest time possible.

In a lucid and approachable way, we have provided the necessary insights on how to blend the two in order to combine their complementary strengths. First and foremost, we have shown that machines, as the means of utmost importance, could probe inexperienced workers' capabilities to build schemas of various categories without supervision or interference from experts in the field. Second, we have illustrated that such collaborative tools can maximize worker efficiency as they could enhance their motivation or inspiration and mitigate any cognitive dulling.   

On a second front, as it is hard to achieve human-level intelligence in machines, these kinds of tools could help the research community focus on discovering new ways to amplify human and machine intelligence. In this sense, we could be looking forward to a slew of applications where people could use automated tools to triangulate solutions for their daily problems. Moreover, as technology, particularly AI, continues to develop, human-computer collaboration platforms could be seen as inevitable stepping tools towards innovation and growth.

\clearpage
\bibliographystyle{splncs04}
\bibliography{isaakJ}
\end{document}